\documentclass{article}
\usepackage{epsfig}
\usepackage[latin1]{inputenc}
\usepackage{xspace}
\usepackage{amssymb}
\usepackage{a4wide}

\def\S12{\mathrm{S}_{1,2}}

\newcommand{\ssu}{$SU(2)_L\times SU(2)_R\times U(1)_{B-L}\,$}

\newcommand{\sulu}{$SU(2)_L\times U(1)_Y$}

\newcommand{\matr}{\left( \begin{array}}
\newcommand{\ematr}{\end{array} \right)}

\newcommand{\dis}{\displaystyle}

\newcommand{\bea}{\begin{eqnarray}}
\newcommand{\eea}{\end{eqnarray}}
\newcommand{\ba}{\begin{array}}
\newcommand{\ea}{\end{array}}
\newcommand{\beqa}{\begin{eqnarray}}
\newcommand{\eeqa}{\end{eqnarray}}

\newcommand{\lsim}{{\;\raise0.3ex\hbox{$<$\kern-0.75em\raise-1.1ex
\hbox{$\sim$}}\;}}
\newcommand{\gsim}{{\;\raise0.3ex\hbox{$>$\kern-0.75em\raise-1.1ex
\hbox{$\sim$}}\;}}

\newcounter{allequation}
\newcommand{\subnumbers}{\setcounter{allequation}{\value{equation}}
      \addtocounter{allequation}{1}
      \setcounter{equation}{0}
       \renewcommand{\theequation}{\arabic{allequation}\alph{equation}}}
\newcommand{\normalnumbers}{\setcounter{equation}{\value{allequation}}
                       \renewcommand{\theequation}{\arabic{equation}}\\}

\begin{document}

\title{
  \begin{flushright}
    {\small CFNUL 2/2000}
  \end{flushright}
  {\sc Searching Scalar Resonances with Ultra-high Energy
  Neutrinos}\\
}
\author{L. Brücher${}^{a,1}$, P. Ker\"anen${}^{a,2}$, J. Maalampi${}^{b,3}$\\[0.5cm]
  {\em \small ${}^a$ Centro de F\'\i sica Nuclear de Universidade de Lisboa,}\\
  {\em \small Av. Prof. Gama Pinto 2, 1649-003 Lisboa, Portugal}\\[0.5cm]
  {\em \small ${}^b$ Department of Physics, Unversity of Helsinki}\\
{\em \small  and}\\
{\em \small $ $ Helsinki Institute of Physics,}\\
  {\em \small P.O. Box 9, FIN-00014 Helsinki, Finland}\\[2.5cm]}
\date{\today} \maketitle

\begin{abstract}

We study the prospects of detecting signals of a resonant
scattering of high-energy cosmic neutrinos on electrons in the
atmosphere. Such a process is possible through an  s-channel
exchange of a isotriplet scalar particle predicted by some
particle physics theories. We estimate the event rates for a
reference detector setup with plausible assumptions on the
interaction strengths and energy resolutions. We find as the most
promising process  the resonance production of tau neutrinos
whose signature would be a "quiet" (in contrast with a hadronic
"bang") production of the tau lepton followed by a more noisy
decay in  downstream.

\end{abstract}
\vspace*{1.5cm}

\begin{flushleft}
  PACS number(s): 12.60.Fr,\ 13.15.+g,\ 13.85.Tp,\ 14.80.Cp,\
  96.40.Pq,\  96.40.Tv,\ 98.70.Sa
\end{flushleft}


\footnotetext[1]{e-mail: bruecher@cii.fc.ul.pt}
\footnotetext[2]{e-mail: keranen@cii.fc.ul.pt}
\footnotetext[3]{e-mail: Jukka.Maalampi@Helsinki.fi}

\thispagestyle{empty}

\newpage



\section{Introduction}

It was suggested long ago by Glashow \cite{Gla1} that one could
look for the effects of the $W$-boson in the spectrum of
 hadronic cascades   produced via a resonant scattering of cosmic antineutrinos with electrons
in the atmosphere.  This suggestion was made before the discovery
of the $W$-boson in laboratory experiments. We will revive
Glashow's idea by  applying it to another resonant process, the
electron-neutrino scattering via an isotriplet scalar exchange.
Isotriplet scalars are predicted by many
models~\cite{tripletreview} and they may play role in the
so-called seesaw mechanism of neutrino masses \cite{seesaw}.
There is no evidence of their existence from laboratory
experiments so far. On the other hand the data still allows their
couplings to leptons to be large, in contrast with the couplings
of the ordinary Higgs scalars~\cite{HuMaPiRa,hui}. We shall study
whether the plans to build large neutrino telescopes and to use
air shower arrays to detect horizontal neutrino-induced air
showers would open a possibility to see signals of these
particles in the cosmic ray spectra. We evaluate the event rates
of the resonant isotriplet mediated electron-neutrino scattering
for plausible isotriplet scalar masses in various detectors and
compare them with the corresponding background rates.


We do not specify the model but assume simply that at the
energies relevant for our studies there exists, in addition to the
Standard Model particles, an isotriplet of scalar fields
$(\dis\Delta^{0},\dis\Delta^{-},\dis\Delta^{--})$ carrying
\sulu\  quantum numbers $(I_3,Y)=(3,-2)$. The couplings of the
triplet $\dis\Delta$ to leptons are governed by the Lagrangian
\begin{equation}
  {\cal{L}}^{\rm Yukawa}_{\Delta} = ih_{\ell\ell'} \Psi_{\ell
    L}^TC\sigma_2\Delta\Psi_{\ell' L}  + h.c.,
\label{yukawa}
\end{equation}
where $\Psi_{\ell L}=(\nu_{\ell L}, \ell_L)$   and
$\ell=e,\mu,\tau$. This interaction breaks the lepton number by
two units. One should note that $\dis\Delta$ does not couple to
quarks, which has important consequences in respect to its possible
astrophysical manifestations.

The interactions (\ref{yukawa}) allow for the $e^-\nu_e$
annihilation in s-channel, whose astrophysical signal we will
consider in the following. The present phenomenological
constraints on the couplings $h_{\ell\ell'}$ are the following
(see e.g. \cite{hui}): \subnumbers\begin{eqnarray}
  \label{limits}
  h_{e\mu}h_{ee} & < &
   3.2\times 10^{-11}\;{\rm GeV}^{-2}\cdot M_{\Delta^{++}} ^2\\
  h_{ee}^2 & \lsim &
   9.7 \times 10^{-6}\; {\rm GeV }^{-2}\cdot M_{\Delta^{++}} ^2\\
  h^2_{\mu\mu} & \lsim &
   2.5\cdot 10^{-5}\; {\rm GeV}^{-2}\cdot M_{\Delta^{++}}^2\\
  h_{ee} h_{\mu\mu } & \lsim &
   5.8 \cdot 10^{-5}\; {\rm GeV}^{-2}\cdot M_{\Delta^{++}}^2\\
  h_{e\mu }h_{\mu\mu} & \lsim & 2 \cdot 10^{-10}\; {\rm GeV}^{-2}\cdot
   M_{\Delta^{++}}^2 \enskip ,
\end{eqnarray}\normalnumbers
where $M_{\Delta^{++}}$ is in GeV. There exist no strict constraints on
the corresponding couplings with $\tau$-leptons and neutrinos.
It is conceivable to assume that the masses of the
all members of the triplet $\Delta$ are more or less equal, and
hence these limits can be taken to be valid also for the couplings of
the singly charged triplet Higgs boson. In the following we will consider
two scalar mass values, 150 GeV and 300 GeV. In the former case the maximum values of the
Yukawa couplings for the first two generations  are
$h_{ee }=0.47$ and $h_{\mu\mu }=0.75$.  In the latter case  the couplings
can have any value up to the vacuum
stability limit of $h_{\ell\ell'}={\cal O}(1)$.


We will consider the process
\begin{equation}
  \nu_e e^- \rightarrow \Delta^- \rightarrow \ell^- \nu_\ell ,\
  (\ell=e,\mu,\tau).
\label{process}\end{equation}
The total cross section of the process in the laboratory frame is
given by
\begin{equation}
  \label{eq:sigma}
  \sigma = \frac{h_{ee}^2h_{\ell\ell}^2}{4\pi}\times
           \frac{m_e E_\nu}{[m_{\Delta^-}^2-2m_eE_\nu]^2 +
           \Gamma_{\Delta^-}^2 m_{\Delta^-}^2 },
\end{equation}
where $E_{\nu}$ is the energy of the cosmic neutrino and $\Gamma_{\Delta^-}$
is the total width of the isotriplet scalar.
 One should  emphasize that the cross
section depends quadratically on $h_{ee}$ and $h_{\ell\ell}$. In
our numerical calculations we assume for these coupling the
largest values allowed by the present laboratory data. If their
values in reality are less than these maximum values, the event
rates presented  below should be rescaled accordingly.

\begin{figure}[htbp]
 \epsfig{file=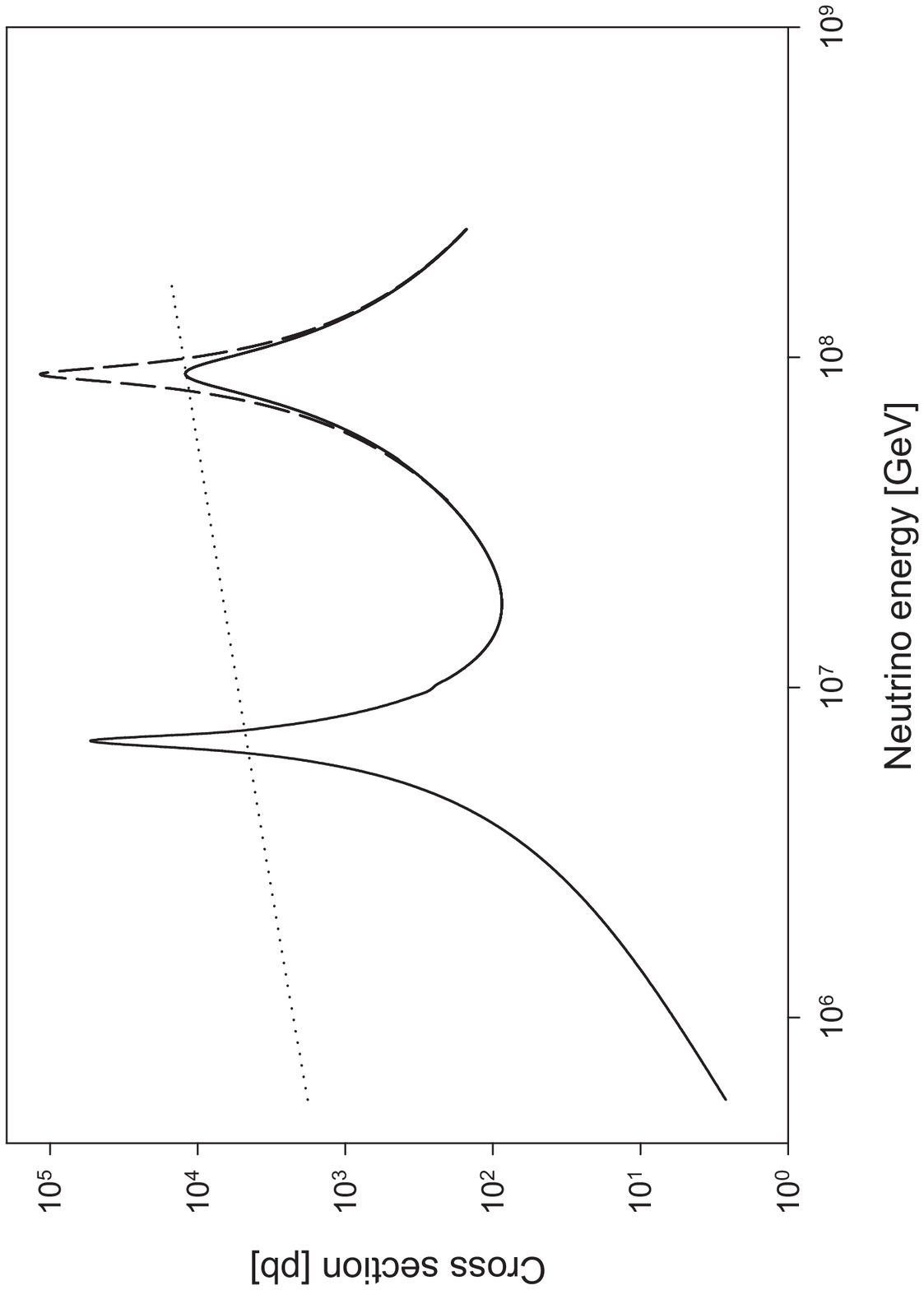,width=8cm,angle=270}
    \caption{Glashow resonance (left peak) and $\Delta$-resonance
      (right peak) with $h_{ee}=h_{\ell\ell}=1$,
      $m_\Delta=300$~GeV and $\Gamma_\Delta=34.5$~GeV (solid line) and with $h_{ee}= 1, h_{\mu}=h_{\tau}=0$,
      $m_\Delta=300$~GeV and $\Gamma_\Delta=11,5$~GeV (dashed line). The straight dotted
      line is the deep inelastic scattering cross section for the
      processes $\nu_\ell N \rightarrow \ell^- X$ and
      $\bar{\nu}_\ell N\rightarrow \ell^+ X$.}
    \label{fig:sigboth}
\end{figure}

Fig.~\ref{fig:sigboth} represents the cross section of the
process~(\ref{process}) for $m_\Delta=300$~GeV and
$\Gamma_\Delta=34.5$~GeV  (right peak) as a function of the cosmic neutrino
energy, in comparison with the cross section of the
Glashow process (left peak). In the same figure we also give the
cross section for the deep inelastic $\nu_\ell$-nucleon and
$\bar{\nu}_\ell$-nucleon scattering processes, which will build
the main background in some detectors. Notice that for all
processes only one neutrino (or antineutrino) flavour is produced;
we assume that one can distinguish between different flavours.
For neutrino energies between $10^{7}\,{\rm GeV}\lesssim
E_\nu\lesssim10^{12}\,{\rm GeV}$ this cross section is given by
\cite{Gand}
\begin{equation}
\sigma_{\nu N+\bar{\nu}N}(E_\nu)\simeq 15.64\times10^{-36}(E_\nu/10^{9}
  \,{\rm eV})^{0.363}\,{\rm cm}^2 \enskip .\label{cross}
\end{equation}
Note that the nucleon to electron ratio is around 2 for most
materials, which will additionally increase this background.

Pion photoproduction processes in the sources
($p\gamma\rightarrow\pi^+n;\ \pi^+\rightarrow\mu^+\nu_\mu;\
\mu^+\rightarrow e^+\nu_e\bar{\nu}_\mu$) suggest that no
$\nu_\tau$ are produced in the sources, and $\nu_\mu$ are
produced twice as many as $\nu_e$. The SuperKamiokande
atmospheric neutrino data indicates that the muon neutrino has a
large mixing with some other neutrino, presumably the tau
neutrino~\cite{atmtheory}. The combined solar neutrino data
favours large mixing angle solutions (LMA, LOW) for the electron
neutrino mixing with an active neutrino~\cite{solar}. This would
lead the flavour composition of the extragalactic neutrino flux
to be even in all flavours~\cite{Pakvasa,LuisJukkaMe},
$N_\tau:N_\mu:N_e=1:1:1$. Note, however, that the discussion of
the solar, atmospheric and laboratory neutrino data is not
settled yet, both three and four neutrino solutions still being
possible~\cite{CarlosLisi}. Nevertheless, assuming the most
plausible neutrino mixing scenario, all three charged leptons are
produced in equal amounts in the deep inelastic scattering
background processes described above.




Let us now move to the possible detection of the
process~(\ref{process}). Currently two major detector designs for
observing high energy neutrinos are discussed, namely atmospheric
detectors and ice or water detectors. These detectors have in
common that they detect fluorescent light emitted by energetic
charged particles. Obviously, in our case the charged particle to
be observed is the high energy lepton $\ell^-$ produced in the
$\Delta^-$ decay. In case $\Delta^-$ decays into $\nu_e$ and
$e^-$, a short electromagnetic cascade will be produced, whose
energy should be measurable. If the energy resolution of the
detector is good enough, the $\Delta^-$-resonance should be
distinguishable from the Glashow resonance. Therefore electrons
are good candidates for detection.  Unfortunately electrons and
positrons are  produced also in deep inelastic scattering
processes that will form a serious background. A final state muon
at these energies propagates a much longer distance than is the
size of any imaginable detector, and furthermore the energy
resolution for muons is much worse than for electrons. Thus they
are not good candidates for the detection of the
process~(\ref{process}).

In contrast, the tau lepton produced by $10^8$~GeV neutrino will
decay typically within 2-3 km from its production point producing a
hadronic shower. The signal is different from the so-called
``double bang event'' produced by $\nu_\tau$-nucleon inelastic
scattering~\cite{Pakvasa}, since there is no particle shower at
the $\tau$ production vertex, i.e.  the first ``bang'' is missing
in our case. If the $\tau$ production occurs outside the
detector, but the decay inside it, the process will be
indistinguishable from the $\nu_\tau$-nucleon deep inelastic
scattering background. Another possible source for background is
the Glashow resonance itself. It can produce high numbers of
$\tau$ (as well as $\mu$ and electron) events, but this is
dominant at lower energies where $\Delta^-$ production is
negligible according to the limits~(2). Thus, with a fair energy
resolution, the Glashow resonance should be distinguishable from
the $\Delta^-$-resonance.

If one can detect the $\tau$ production vertex without a hadronic
shower ("bangless" vertex) and $\tau$ decay after the decay
length of the order of some kilometers, one may measure the
lepton energy and hence also estimate  the neutrino energy. In
that case the most serious background for the processes
$\Delta^-\rightarrow\tau\nu_\tau$ are muon neutrino events and
cosmic ray muons, with a catastrophic energy loss of the muon
after around the same flight distance like $\tau$. The energy
loss may be e.g. due to hitting a nucleus in the atmosphere,
mimicking the $\tau$ decay. The difference in the signature of
muon catastrophic energy loss should be simulated for different
detectors, but it is beyond the scope of this letter.
Nevertheless, one should be able to statistically distinguish the
big sample of decaying $\tau$-leptons, having definite energy and
distance from the production vertex, compared to the randomly
distributed muon events.

In the following event rate calculations we use as our reference
the active galactic nuclei (AGN) models described in
~\cite{proth}. The background calculations are performed in ~\cite{Gand}.
None of these AGN models have been tested with actual data.
Nevertheless, we emphasize the potential importance of observing
the high energy neutrino flux and looking at possible resonance
effects, which are independent of  neutrino sources or the
astrophysical models describing neutrino production.



  We are interested here in
close to horizontally incoming particles. The event rate of the
horizontal air showers in a neutrino detector with an area $A$ is
 \begin{equation}
   \label{eq:rate}
   Rate =  A \times t \times \int dE_\nu \frac{dN}{dE_\nu} \times
    \sigma (E_\nu) \rho_e(h) dh \enskip ,
 \end{equation}
where $t$ denotes the time
and $\rho_e(h)$ is the electron density, calculated from the
American Standard Atmosphere~\cite{ASA}, depending on the height
$h$ from the surface of the Earth. The yearly event rate per
steradian is shown in table~\ref{tab:at}. We have done our
estimates for each different AGN model with two different energy
resolutions of the detector with a surface area of
$10^6$~km$^2$. The first column corresponds to the
energy resolution of  25~\% (or $2.5\times 10^7$~GeV) and the
second column that of  5~\% (or $5\times 10^6$~GeV). We consider
the first energy resolution plausible, while the second one is
probably optimistic~\cite{Zas}. The event rates have been
calculated in each case for two situations. On the first line it
is assumed that all three neutrinos have the coupling strenght of
$h_{ee}=h_{\mu\mu}=h_{\tau\tau}=1$, corresponding to the decay
width $\Gamma(\nu_e e^-\rightarrow \nu_\ell \ell^-) = 34.5$ GeV,
while on the second line it is assumed that $h_{ee}=1$ and
$h_{\mu\mu}\simeq h_{\tau\tau}\simeq 0$, corresponding to the
width 11.5 GeV.  The increase of events in the case of low energy
resolution and large width is perhaps hard to detect. In the case
when only the electron neutrino has a large coupling with
$\Delta^-$ one has better chances to see the increase. The best
channel for detecting the effect may be
$\Delta^-\rightarrow\tau\nu_\tau$. If one can distinguish the
production vertex of $\tau$ to be without hadronic shower
(without the first bang), there will be plenty of events per year
to be seen.
\begin{table}[htbp]
  \begin{center}
    \begin{tabular}{c|rr|rr|rr}
     $\Gamma_\Delta$ [GeV] & \multicolumn{2}{c|}{AGN P96} &
       \multicolumn{2}{c|}{AGN SS91} &  \multicolumn{2}{c}{AGN M95} \\
     \hline
     \hline
    34.5 &  606 &  230 &  207 &  75 &  229 &  88 \\
     11.5 &  2230 & 1484 & 740 & 482 & 847 & 566 \\
  \hline
   $\nu N+\bar{\nu}N$ & 4829 & 919 & 1793 & 300 & 1801 & 350 \\
  \hline
  \hline
    \end{tabular}
    \caption{Yearly event rate in an atmospheric detector
    with a surface area of $10^6$~km$^2$ for energy intervals
       of $5\times 10^{7}$GeV (left row) and $1\times 10^{7}$GeV
       (right row) for $m_\Delta=300$~GeV.}
    \label{tab:at}
  \end{center}
\end{table}


Water and ice detectors are build in deep water~\cite{Nestor} or
in antarctic permanent ice~\cite{Amanda} and most probably have a
final size of just 1 km${}^3$. Thus, assuming $m_\Delta\gsim
200$~GeV, the events with $\tau$-leptons in the final state are no
longer fully contained in the detector. Consequently the observed
$\tau$'s decaying inside the detector will no longer be
distinguishable from $\nu_\tau$-nucleon deep inelastic
scatterings. Signatures of the muons or $\tau$-leptons that are
produced inside the detector but that leave it are probably as
well inseparable from each other. Events with electrons in the
final state might give a signal with long detection times. The
event rates per year under these conditions are shown in
table~\ref{tab:ice}.\footnote{ Note that
  $\rho_{e(ice)} = 3\times 10^{29} m^{-3}$ and that the detector can
  work almost 100\% a year. Moreover we assume that a total steradian
  of $2\pi$ can be covered.}
\begin{table}[htbp]
  \begin{center}
    \begin{tabular}{c|rr|rr|rr}
     $\Gamma_\Delta$ [GeV] & \multicolumn{2}{c|}{AGN P96} &
       \multicolumn{2}{c|}{AGN SS91} &  \multicolumn{2}{c}{AGN M95} \\
     \hline
     \hline
    34.5 &  4.0 &  1.4 &  1.3 &  0.5 &  1.4 &  0.5 \\
     11.5 &  14 & 9 & 4.6 & 3.0 & 5.3 & 3.5 \\
  \hline
   $\nu N+\bar{\nu}N$ & 30 & 6 & 11 & 2 & 11 & 2 \\
  \hline
  \hline
    \end{tabular}
    \caption{Events in a cubic kilometer sized
    ice/water detector for energy intervals
       of $5\times 10^{7}$GeV (left row) and $1\times 10^{7}$GeV
       (right row) for $m_\Delta=300$~GeV.}
    \label{tab:ice}
  \end{center}
\end{table}

Water and ice based neutrino telescopes are effective if $\tau$
decays inside the detector. Observing fully contained events, we
allow a maximum path of 500 m and additional 250 m for the
observation of the $\tau$-decay. In this case the resonant
production of $\Delta^-$ is supposed to happen while the
ultra-high energy neutrino propagates the first 250 m path inside
the detector. The mass range $m_\Delta\lsim 150$~GeV can be tested
this way in cubic kilometer detectors.
Event rates for this case are presented in table~\ref{tab:ice2}
for $h_{ee}=0.47$, $h_{\mu\mu}=0.75$ and $h_{\tau\tau}=1$, which
are the largest values allowed  by the present bounds~(2).
Non-observation of $\tau$ events with these rates would set new
constraints on the couplings.
\begin{table}[htbp]
  \begin{center}
    \begin{tabular}{c|r|r|r}
      & AGN P96 & AGN SS91 & AGN M95 \\
     \hline
     \hline
    $\Delta^-$-resonance &  8.1 &  16.0 &  1.5 \\
     Glashow-resonance &  14.1 & 70.5 & 1.2 \\
  \hline
   $\nu N+\bar{\nu}N$ & 10.7 & 21.5 & 1.9 \\
  \hline
  \hline
    \end{tabular}
    \caption{Events in a cubic kilometer sized ice/water detector for an energy interval
       of $1\times 10^{7}$GeV for $m_\Delta=150$~GeV and $\Gamma_\Delta=6.2$~GeV.
In this case it has been assumed that $h_{ee}=0.47$,
$h_{\mu\mu}=0.75$ and $h_{\tau\tau}=1$.}
    \label{tab:ice2}
  \end{center}
\end{table}


In conclusion, we have investigated the prospects of observing the
possible resonant scattering of high-energy cosmic neutrinos on
electrons in the atmosphere via an isotriplet scalar exchange. It
turn out that  it may be, even with large lepton-scalar
couplings, hard to distinguish events with muons or electrons in
the final state. On the other hand, events with $\tau$-leptons in
the final state have an almost backgroundless signature provided
it can be checked for each event that there is no hadronic shower
in the production vertex of the $\tau$. In large atmospheric
arrays, of the size of the OWL~\cite{owl}, one can expect to see
of the order of a thousand such events per year.
%

\bigskip

{\it Acknowledgements.} We would like to thank Robert Stokstad and
Enrique Zas for helpful discussions. J.M. wishes to thank the Centro de F\'\i
sica Nuclear de Universidade de Lisboa for hospitality. P.K. wishes to thank the
Lawrence-Berkeley National Laboratory for hospitality during the final
stages of this work and the
 Jenny and Antti Wihuri foundation and the Finnish
Academy of Sciences and Letters for grants. L.B. and P.K. are
supported by JNICT under contracts No. BPD.16372 and No.
BPD.20182. This work has been supported by the Academy of Finland under the project no. 40677.





\end{document}